\newcommand{\bi}{\bibitem}
\newcommand{\be}{\begin{eqnarray}}
\newcommand{\ee}{\end{eqnarray}}
\newcommand{\rar}{\rightarrow}
\newcommand{\lar}{\leftarrow}

\newcommand{\ds}{\partial \!  \! \! /}

\documentstyle[12pt,lathuile]{article}
\begin{document}
\title{ 
ANTIMATTER IN THE UNIVERSE
}
\author{
A.D. Dolgov\\
{\em 
INFN, Ferrara, Italy}\\  {\em and}\\ 
{\em ITEP, Moscow, Russia
} \\
}
\maketitle
\baselineskip=14.5pt
\begin{abstract}

Different scenarios of baryogenesis are briefly reviewed from the
point of view of possibility of generation of cosmologically 
interesting amount of antimatter. It is argued that creation of
antimatter is possible and natural in many models. In some models
not only anti-helium may be produced but also a heavier 
anti-elements and future observations of the latter would be critical
for discovery or establishing stronger upper limits on existence of
antimatter. Incidentally a recent observation of iron-rich quasar
may present a support to one special model of antimatter creation.

\end{abstract}
\baselineskip=17pt
\newpage
\section{Introduction\label{sec-intr} }
Our region of the universe is certainly dominated by matter, protons,
electrons, and nuclei consisting of protons and neutrons. No antimatter
in any significant amount is observed. A little of antiprotons and
positrons in cosmic rays can be explained by their secondary origin
in collisions of protons, electrons or photons with usual matter.
Cosmological excess of matter over antimatter is described by the
ratio
\be
\beta = {N_B - N_{\bar B} \over N_\gamma} \approx 6\cdot 10^{-10}
\label{beta}
\ee
where $N_{B,\bar B,\gamma}$ are respectively
the cosmic number densities of baryons,
antibaryons, and photons in microwave background radiation (CMBR).
At the present day $N_\gamma = 412/ {\rm cm}^3$ and $N_B \gg N_{\bar B}$ 
(at least in our  neighborhood.). It is believed that in the early 
universe, at high temperatures, $T> 100$ MeV, the number densities of
baryons and antibaryons (at these temperatures they were in quark
state) were almost equal with relative accuracy of the order of
$\beta$. 

According to simple models of baryogenesis, pioneered by
Sakharov\cite{sakharov67} in 1967, baryon asymmetry is homogeneous, 
i.e. $\beta$ does not depend on space points, and the total baryonic 
charge of the universe is non-zero:
\be
B_{tot} = \int \beta\,d^3 x \neq 0
\label{Btot}
\ee
Still it is not excluded neither theoretically nor observationally
that this may be not so and we face the following big questions:

\begin{enumerate}

\item{}
Is $\beta = const$ or it could be a function of space point,
$\beta = \beta (x)$?

\item{}
If $\beta = \beta (x)$ what is the characteristic scale $L_B$ of its
variation? Especially interesting is if $L_B$ may be smaller than the
present-day horizon $L_{hor} \sim 3 $ Gpc, or it is possible that
$L_B < L_{hor}$?

\item{}
If $\beta $ indeed varies, may it be that in some astronomically
sizable regions $\beta <0$, that is some parts of the universe are
antimatter dominated? 

\item{}
If $\beta <0$ is allowed what is the global baryonic charge of the
universe? Is $B_{tot} \neq 0$, so the universe is globally charge
asymmetric or $B_{tot} =0$ and the universe is globally charge
symmetric.  

\end{enumerate}

In this talk I will discuss the present observational bounds on
existence of antimatter and scenarios of baryogenesis that might lead
to astronomically interesting antimatter domains or antimatter objects. 

\section{Observational limits \label{sec-lim}}

Antimatter may be observed by the energetic gamma rays produced by 
$p \bar p$-annihilation in the regions where matter and antimatter
domains are in contact. According to the analysis made in the 70th,
the level of gamma ray flux in the 100 MeV energy range demands that
the nearest anti-galaxy should be at the distance larger than 
10-15 Mpc, as reviewed in ref.\cite{steigman76}. A discussion of 
possible search for cosmological antimatter can be found in 
refs.\cite{chechetkin82,stecker85}.

More recently the bound quoted above was strongly improved in 
ref.\cite{cohen98} up to Gigaparsec range. It was argued there that 
the matter-antimatter
domains could not be too far separated because, otherwise, density
deficit in baryon poor regions between the domains would be
noticeable by angular fluctuations of CMBR. The minimum observable
scale is about 20 Mpc. Moreover, below approximately this scale 
fluctuations are strongly smoothed down by photon
diffusion\cite{silk68}. Fluctuations at smaller scales may escape
observations in angular spectrum of CMBR but in this case
proton/antiproton diffusion would bring them into contact at later
stage and there would be a burst of annihilation which might be
observable in cosmic gamma rays. One would naively expect that the
annihilation, when starts, would produce excessive pressure in the the
region of annihilation which would push matter-antimatter domains
apart. Hence the efficiency of the process would be low and large
antimatter domains in our neighborhood would be allowed. 
In fact, as shown in ref.\cite{cohen98}, the situation is opposite:
energy and extra pressure produced by the annihilation would be
released far away from the annihilation region because of a large
mean-free path of the annihilation products. As a result matter and
antimatter would be pushed towards stronger contact and the
annihilation would be more and more efficient. This mechanism allowed
to obtain a very strong bound quoted above. 

As mentioned by the authors\cite{cohen98} 
the bound is valid in the case of 
baryo-symmetric cosmology when matter and antimatter are equally
abundant in the universe and for the case of adiabatic density
perturbations. For isocurvature perturbations the behavior could be
different. Initially density contrast was absent, so the energy
densities of baryonic and antibaryonic domains and the baryon-poor
boundary between them were the same. When baryons became
non-relativistic energy density of the regions with larger baryon (or
antibaryon) numbers became larger than in baryon-poor regions. On the
other hand, the temperature of photons in baryon-poor regions became
higher than in (anti)baryon rich regions because nonrelativistic
matter cools faster. Higher photon temperature
would lead to an excessive pressure which would put baryons and 
antibaryons apart diminishing probability of annihilation.
This may allow antibaryonic domains to be much closer to us than
Gigaparsec, especially if the universe is not baryo-symmetric and
the amount of cosmic antimatter is noticeably smaller than amount of
matter. Unfortunately qualitative results for this case are not yet 
available. 

\begin{figure}[t]
 \vspace{9.0cm}
\includegraphics{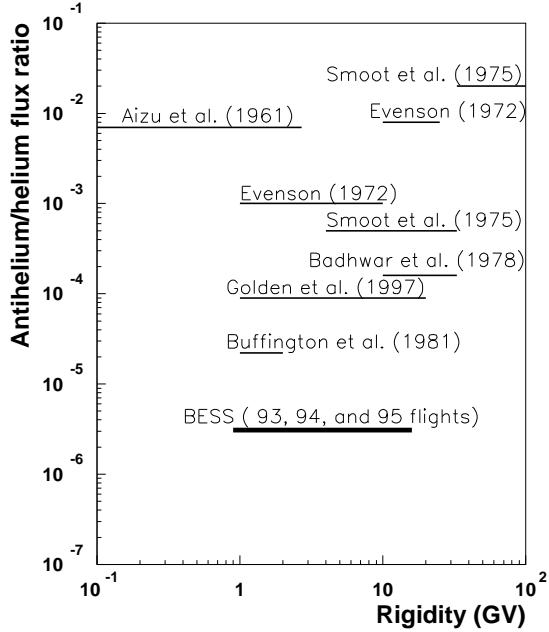}
 \caption{\it
The upper limit of $\bar He/He$ flux ratio of 
ref.\cite{bess97} together with previous limits.
    \label{fig-he} }
\end{figure}

An unambiguous proof of existence of cosmic antimatter would be
observation of anti-nuclei in cosmic rays. An observation even of a
single $^4 \bar He$-nuclei or a heavier one would demonstrate that 
primordial antimatter indeed exists not too far from us. The
present-day upper limits are not too restrictive. They are summarized
in ref.\cite{bess97} and presented in Fig.~\ref{fig-he}.

\begin{figure}[t]
 \vspace{9.0cm}
\includegraphics{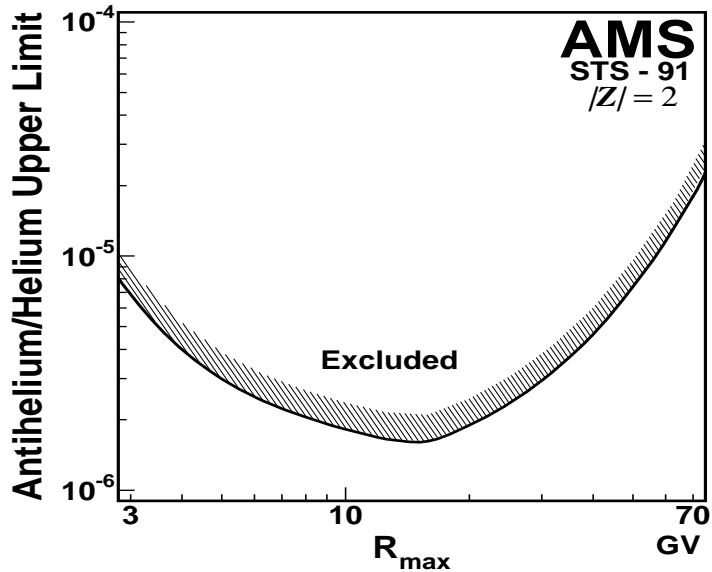}
 \caption{\it
Upper limits on the relative flux of anti-helium to helium,
            at the 95\% confidence level\cite{ams99}.
            These results are independent of the incident 
anti-helium spectra.
    \label{fig-ams} }
\end{figure}

A stronger limit, though rapidity dependent, is presented in 
ref.\cite{ams99}, see Fig.~\ref{fig-ams}. Future AMS mission on
International Space Station may improve the bound by three orders of 
magnitude.

\section{Generation of cosmological baryon asymmetry: general features
\label{sec-gen}}

The mechanism of dynamical creation of excess of particles over
antiparticles in the early universe was suggested in 1967 by
Sakharov\cite{sakharov67} and is accepted now as one of the
cornerstones of the modern cosmology. The well known three principles
of baryogenesis are:
\begin{enumerate}
\item{}
Non-conservation of baryonic charge.
\item{}
Breaking of symmetry between particles and antiparticles (violation
of C and CP invariance).
\item{}
Deviation from thermal equilibrium in the primeval plasma.
\end{enumerate}

In fact none of these conditions are absolutely obligatory (see
discussion in the review\cite{dolgov92} and in what follows) 
but in simple models they all should be fulfilled.

At the time when the model of baryogenesis was suggested the 
hypothesis of baryon non-conservation was probably the weakest one.
40 years ago the common belief was : ``we exist, so proton must 
be stable''. Now the point of view is exactly opposite: ``we exist, 
so proton must be unstable''. Justification for the latter is that 
the universe with suitable for life conditions might emerge only if
baryons were not conserved. In particular, inflation would not be
possible if baryonic charge was conserved. So at the present day
existence of the universe is the strongest ``experimental'' fact in
favor of baryon non-conservation. Theory also evolved in the 
favorable direction. Models of unification of strong and electro-weak 
interactions predict non-conservation of 
baryons\cite{pati74,georgi74}. Moreover it was shown that even the
standard electro-weak theory, which respects baryon conservation in
the Lagrangian, breaks this law by quantum corrections\cite{thooft76}.

Deviation from thermal equilibrium in primeval plasma, stressed in 
ref.\cite{okun76}, is usually small but non-vanishing for massive
particles. It is roughly equal to:
\be
{\delta f / f} \sim \left( {H /\Gamma} \right)
\left( {m / T} \right)
\label{df/f}
\ee
where $f$ is the distribution function of the massive particles with
mass $m$, $\delta f$ is its deviation from equilibrium, $\Gamma$ 
is the characteristic rate of the reaction with particles in question,
and $H$ is the Hubble parameter. One sees that for $m \ll T$ the 
deviation can be very big but for low temperatures the number density
of massive particles is Boltzmann suppressed and the net effect is
small. So usually the most favorable period for generation of
asymmetry is when $T \sim m$. 
A large deviation from thermal equilibrium might also take place in the
case of strongly first order phase transition with large supercooling
and coexistence of two phases.


Breaking of symmetry between particles and antiparticles is
established at experiment but theory of the effect is still
uncertain. There are many possible theoretical models but we do not
yet know which mechanism or several different ones is/are realized.

The simplest and generally accepted is explicit C(CP)-violation which 
is realized by introducing complex coupling constants or masses into
fundamental Lagrangian. In the models of baryogensis based on this
assumption cosmological  baryon asymmetry has a definite sign
determined by the particle physics and usually $\beta$ is space point
independent. 

However charge symmetry may be broken spontaneously as suggested in
ref.\cite{lee73}. In such a model vacuum state is degenerate
and a complex scalar field acquires different vacuum expectation 
values corresponding to different vacuum states. 
The sign of C(CP)-violation is
different there and baryogenesis would end up either with baryons or
antibaryons. It was argued in ref.\cite{brown79} that if  this type 
of charge symmetry breaking is realized in nature the universe would
be globally charge symmetric with chaotically distributed baryonic and
antibaryonic domains. However the size of the domains happened to be
too small and some moderate period of exponential expansion was
necessary\cite{sato81} to make the model consistent with the data. 
Still the model encounters serious problems. First, there must be
domain walls with huge energy density separating different vacuum
states. Existence of such walls contradicts the observed homogeneity 
and isotropy of the universe\cite{zeldovich74} and one should invoke a
mechanism to avoid formation of such walls or to destroy them at a
later stage. Second, the model predicts charge symmetric universe with
close contact of matter-antimatter domains and according to
ref.\cite{cohen98} discussed above the size of domains should be close
or larger than horizon to avoid contradiction with the observed gamma
ray background. One can however ``create'' charge asymmetric universe
dominated by baryons but with some antibaryonic domains if both explicit
and spontaneous mechanisms of C(CP) violation are 
operative\cite{chechetkin82z}.

One more mechanism of breaking the symmetry between particles and
antiparticles which might be effective only in the early universe 
can be called stochastic\cite{dolgov92}. We believe that there exist
many complex scalar fields with the mass smaller than the Hubble 
parameter at inflation; the latter could be as large as 
$10^{-5} m_{Pl} \approx 10^{14} $ GeV. Such ``light'' fields are
infrared unstable in De Sitter space-time\cite{bunch78} and because
of that they acquire non-zero vacuum value
\be
\langle \phi^2 \rangle ={ 3 H^4\over 8 \pi^2 m^2}
\label{phi2}
\ee
The field tends to this asymptotic value as 
$\langle \phi^2 \rangle = H^3 t/(2\pi)^2$\cite{vilenkin82}.
Thus the field could be displaced from (mechanical) equilibrium by 
quantum fluctuations during inflation and if it did not relax to
equilibrium before baryogenesis its non-zero amplitude acted as
the vacuum condensate of the field that induced spontaneous
breaking of charge symmetry in the model discussed above. An important
difference with respect to the model of spontaneous C(CP)-breaking is
that the field $\phi$ would ultimately relax down to equilibrium 
point $\phi =0$ and no cosmic domain walls would remain.
Similar type of C(CP)-violation is present in some models of
baryogenesis described below.

Since stochastic C(CP) breaking would not survive to the present day
the observed CP-violation in particle physics should be prescribed to
another mechanism, e.g. to the explicit one. Correspondingly, if both
mechanisms are operating and if the amplitude of the stochastic one is
larger than the explicit, there could be domains of antimatter in the
universe but their fraction would be smaller than 50\%. In this 
scenario large isocurvature fluctuations can be expected.

\section{Models of baryogenesis \label{sec-bar}}

\subsection{Heavy particle decays}

It is historically first model of 
baryogenesis\cite{sakharov67,kuzmin70} which later received robust
theoretical foundation based on GUTs - grand unified 
theories\cite{weinberg79} (for more references and development
see e.g. 
reviews\cite{dolgov81,kolb83,dolgov92,dolgov98,riotto99}).
The mechanism is quite simple: if there are GUT heavy bosons, $X$, 
out of thermal equilibrium, then e.g. the decays $X\rar 2q $ and 
$\bar X\rar 2\bar q$, where $q$ is a quark, may have different 
probabilities due to C(CP)-breaking and as a result an excess of
baryons over antibaryons may be created by these decays. In 
ref.\cite{kuzmin70} the decays of Majorana fermion were considered,
so the same particle may decay into charge conjugated channels
with different branching ratios.

If GUT scale is about $10^{16}$ GeV, as indicated by the recent
data, then the deviation from equilibrium at $T\sim m_X$ is large
(see eq.~\ref{df/f}) and the model is capable to supply the necessary
excess of baryons over antibaryons. However it is questionable if the
universe ever reached temperatures about $10^{16}$ GeV and if such 
heavy bosons were abundantly produced. 

In the standard versions the model does not lead to creation of
antimatter, the asymmetry $\beta$ is determined by elementary 
particle physics and $\beta$ is predicted to be a universal constant
over all the universe. To avoid this conclusion one needs to include 
additional fields and interactions which are not
present in the usual GUTs.

\subsection{Electroweak baryogenesis}

The standard electroweak (EW) theory possesses all the features 
necessary for baryogenesis: breaking of symmetry between particles 
and antiparticles, non-conservation of baryonic charge (by chiral 
anomaly) and may lead to a strong deviation from thermal
equilibrium. Breaking of equilibrium due to particle masses 
is very weak, according to eq.~(\ref{df/f}), but the cosmological
phase transition from EW-unbroken to EW-broken phase might be first
order and, if this was the case, thermal equilibrium could be strongly
broken. 

The possibility to create baryon asymmetry of the universe in 
frameworks of known physics makes the model very attractive and
after the pioneering paper\cite{kuzmin85} the model became the most
popular one. However with strengthening of the lower bound on the
Higgs boson mass presented by LEP the first order phase transition 
(which could take place only for sufficiently light Higgs) became 
less and less probable and now electro-weak baryogenesis has lost a 
considerable part of its attraction. For more detail one can see the
reviews\cite{riotto99,rev-ew}.

Creation of antimatter in electroweak scenario was not discussed
but seemingly the situation in simplest versions of the model is
similar to the GUT case: the baryon asymmetry is positive and 
uniform.

\subsection{Baryo-thru-lepto-genesis}

This scenario was suggested in ref.\cite{fukugita86} and
combines the ideas of the two discussed above. First, a lepton
asymmetry is generated in decays of heavy Majorana neutrino,
$\nu_M$,
with the mass about $m_M\sim 10^{10}$ GeV. Later, electroweak 
processes which conserve the difference of baryonic and leptonic 
charges, $(B-L)$, would equilibrate them and as a result there would 
be generated baryon asymmetry equal to a fraction of the initially
produced lepton asymmetry. The necessity of noticeable deviation from
thermal equilibrium demands $\nu_M$ to have very weak interactions.
On the other hand, since the asymmetry could be produced by 
$\nu_M$ decays only in the second order in the L-non-conserving
interaction (see e.g. discussion in ref.\cite{dolgov81}), it cannot 
be large. Subsequent entropy dilution could bring the baryon 
asymmetry down to unacceptably low value. A more optimistic point of
view shared by majority working in the field is that 
baryo-thru-lepto scenario can supply the necessary amount of baryons 
in the universe. For a recent review see e.g. ref.\cite{buchmuller02}.
As for antimatter production, this approach is in the same bad shape
as the other two described above. 

\subsection{Black hole evaporation}

An excess of baryons over antibaryons could be produced by the 
evaporation of low mass black 
holes\cite{hawking75,carr76,zeldovich76}. A concrete mechanism
was suggested in the paper\cite{zeldovich76} and the calculations 
of the effect were performed in ref.\cite{dolgov80}. In the process
of evaporation all the particles with the mass smaller than
black hole temperature, $T_{BH}= m_{Pl}^2 /(8\pi M_{BH})$ can be
produced. A massive meson, still in gravitational field of
black hole could decay into a light baryon and heavy antibaryon
(e.g. $u$ and $\bar t$ quarks) or vice versa. 
The decay probabilities may
be different because of C(CP) violation. Since back capture
of heavy particles by the black hole is more probable that of
light ones, such process could create a net flux of baryonic 
charge into external space, while equal antibaryonic charge
would be hidden inside disappearing black hole. This mechanism 
could explain the observed value of the baryon asymmetry of the
universe if at some early stage the total cosmological energy density
was dominated by those black holes.

Evaporating black holes may not disappear completely. The process 
may stop when their mass drops down to the Planck value. In this 
case such stable Planck mass remnants would contribute into 
cosmological dark matter (see e.g.\cite{dolgov00}). (In the case 
of theories with large extra dimension the mass may be as small as 
TeV.)

Without special efforts this model is also not good for creation
of cosmologically significant amount of antimatter.

\subsection{Spontaneous baryogenesis}

The model was proposed in ref.\cite{cohen87} and is based on the
assumption that $U(1)$-symmetry, related to baryonic or some other
non-orthogonal charge, is spontaneously broken. A toy-model
Lagrangian can be written as:
\be
{\cal L} = - |\partial \phi |^2 +
\lambda \left( |\phi|^2 - f^2 \right) + 
\sum_a \bar \psi_a \left(i\ds + m_a  \right) \psi +
\sum_{a,b} \left(g_{ab} \phi \bar \psi_b  \psi_a + h.c. \right)
\label{unbrkn}
\ee
where some fermionic fields $\psi_b$ possess non-zero baryonic charge,
while some other do not. The theory is invariant with respect to 
simultaneous phase rotation: 
\be 
\phi \lar \phi \exp (i\theta)\,\,\, {\rm and}\,\,\,
\psi_b \rar \psi_b \exp (i\theta) 
\label{rot}
\ee
which ensures conservation of ``baryonic'' charge. In the broken
symmetry phase where $|\phi| = f$ the conservation of baryonic
current of fermions also breaks down (due to presence of the last
term in the Lagrangian above) and the Lagrangian takes the form:
\be
{\cal L} = - f^2 \left( \partial \theta \right)^2 + 
\partial_\mu \theta\, \bar \psi_b \gamma_\mu \psi_b + ...
\label{broken}
\ee
where $\theta$ is the massless Goldstone field induced by the breaking
of the global $U_b$-symmetry. If there are some additional terms
in the Lagrangian producing an explicit symmetry breaking then
$\theta$ would be massive and is called pseudo-Goldstone field. 
Baryogenesis would be much more efficient in the latter case. 

In the homogeneous case when $\theta = \theta (t)$ the second term
in expression~(\ref{broken}) looks like $\dot \theta n_b$ where
$n_b$ is the baryonic charge density. So it tempting to identify
$\dot \theta$ with chemical potential of baryons\cite{cohen87}.
Though it is not exactly so\cite{dolgov95}, still this term shifts
equality between number densities of quarks and antiquarks even 
in thermal equilibrium. 

The sign of the created baryon asymmetry is determined by the
sign of the $\dot\theta$ and could be both positive or negative
producing baryons or antibaryons. To create the matter/antimatter
domain of astronomically large size the $U(1)$-symmetry should be
broken during inflation and in this case the sign of $\dot\theta$
would be determined by the chaotic quantum fluctuations at
inflationary stage. The analysis of density perturbations created
by fluctuating field $\theta$ was done in ref.\cite{turner89}.

In this scenario C(CP)-violation is not necessary
for generation of baryon asymmetry. As a whole the universe would be
charge symmetric. We know however that an explicit C(CP)-violation
is also present. If it also participate in creation of baryon
asymmetry, then the amount of matter and antimatter in the universe 
would be different with unknown ratio that should be determined from 
observations.

\subsection{SUSY condensate baryogenesis \label{ssec-susy}}

If supersymmetry (broken of course) exits in nature then together
with baryon-fermions there should be baryon-scalars. A scalar
field, $\chi$, with non-zero and non-conserved
baryonic charge could develop vacuum
condensate during inflation according to the mechanism discussed
at the end of Sec.~\ref{sec-gen} if its mass is smaller than the 
Hubble parameter at inflation. After the end of inflation this
condensate can decay liberating the accumulated baryonic charge into
usual baryons (quarks). This is the essence of the model, proposed
in ref.\cite{affleck85}, which can be very efficient 
for generation of baryon asymmetry. All features specified above
are typical for SUSY models. In particular, the self-potential
of $\chi$ possesses the so called flat directions (or valleys)
along which the potential energy of the field does not rise.

As a toy model let us consider the potential energy of the form:
\be
U(\chi) = \lambda \left[ |\chi|^4 - \left( \chi^4 + \chi^{*4}
\right)/2 \right] = \lambda |\chi|^4 \left( 1 - 4\cos \theta
\right) 
\label{uofchi}
\ee  
where $\chi = |\chi| \exp (i\theta)$. This potential has four 
valleys $\theta = \pi n/2$ with $n=0,1,2,3$ and is not
invariant with respect to rotation in two-dimensional
$({\rm Re}\chi,\,\,{\rm Im}\chi)$-plane. This leads to
non-conservation of baryonic current of $\chi$. The latter is
defined as
\be
J_\mu^B (\chi) =(-i/2)\left( \chi^* \partial_\mu \chi -
\partial_\mu \chi^* \chi \right) = \partial_\mu \theta\,|\chi|^2 
\label{JchiB}
\ee
In the homogeneous case when $\theta = \theta (t)$ only time
component of the current (i.e. baryonic charge density) is
non-vanishing. There is a very convenient mechanical analogy in
this case. The equation of motion for $\chi$
\be
\ddot \chi + 3 H\dot \chi + U' (\chi) = 0
\label{ddotchi}
\ee
describes classical mechanical motion of point-like particle in
the potential $U(\chi)$. The second term induced by the cosmological
expansion presents liquid friction force. In this language the
baryonic charge of $\chi$ is simply angular momentum of the particle
in this potential. If the potential is spherically symmetric (in two
dimensions) then baryonic charge is conserved, otherwise is not.

The field $\chi$ should generically possess a non-zero mass. It could
be either produced by some soft symmetry breaking after inflation 
was over or might be non-zero even at inflationary stage due to 
explicit mass term in the Lagrangian, $m^2 |\chi|^2$. 
During inflation the ``particle'' was deep along the valley and
when inflation stopped and the Hubble parameter became smaller than
$m$ it starts to roll down to the origin along the potential slope.
The ``particle'' may be displaced a little from the minimum in the
valley and hence some orthogonal oscillations between the walls of 
the valley would be superimposed on its motion down. These 
fluctuations were damped by the Hubble friction and, what's more 
important, by the particle production by the time dependent
field $\chi (t)$. The average value of the angular momentum at this 
stage was zero and no baryonic charge was produced by the decay
of $\chi (t)$. However, when $\chi$ comes closer to the origin
the potential becomes dominated by the spherically symmetric mass
term and the oscillating behavior would change into rotation
around the origin. This corresponds to non-zero 
baryonic charge which all would be transferred to the light particles
produced by the decay of $\chi$. (We assume that interaction of 
$\chi$ with light particles conserves baryonic charge.) 

It is evident that the sign of the baryon asymmetry is determined by
the direction of the rotation, which could be either clock-wise or
anti-clock-wise depending upon chaotic initial conditions. The latter
introduce also an initial breaking of charge asymmetry, so no
C(CP)-violation is necessary as in the example considered in the
previous subsection. 

The asymmetry generated according to this scenario can be very small
because the orthogonal motion (which carries a non-zero angular
momentum) might be strongly damped by the particle 
production\cite{dolgov91} (see also\cite{dolgov92}). The frequency 
of the oscillations is determined by the slope of the potential in 
the direction orthogonal to the valley, 
$m_{eff} = \sqrt \lambda \chi$, and could be large for large 
displacement from the origin. Hence particle production could be
very strong and the baryon asymmetry originated from the orthogonal
motion would be small.

This conclusion can be avoided if there is an explicit C(CP)-breaking
in the theory. One can introduce it assuming that there is a 
non-zero relative phase, $\alpha$, between the coupling constant 
$\lambda$ and mass $m$. The potential of $\chi$ in this case have 
the form:
\be
U(\chi) = \lambda |\chi|^4 \left( 1 - \cos 4\theta \right)
+m^2 |\chi|^2 \left[ 1 - \cos \left( 2\theta + 2\alpha \right)\right]
\label{uodd}
\ee
The evolution of field $\chi$ in this potential would proceed as
follows. Let us first consider the case of $\alpha =0$. If the
field $\chi$ was inside one of the valleys $\theta = 0,\pi$ then
these valleys coincide with mass valleys and the field
would evolve down to zero along these lines and no baryon asymmetry
would be produced. The picture would be very much different if the
field was initially in one of the valleys with $\theta = \pm\pi/2$
which are orthogonal to mass valleys. When the field
approaches the origin and the mass term in the potential becomes
dominating, the motion along the line $\theta = \pm\pi/2$ would be
unstable and 
the field would start to move in direction of one or
other mass valleys acquiring a non-zero (and possibly large)
angular momentum. The choice between clock-wise or anti-clock-wise
directions would be chaotic and the universe would equally consist 
of baryonic and antibaryonic domains and would be globally
baryo-symmetric. The size of the domains would be astronomically
large if the trend to mass valleys would begin still at inflation. 

If $\alpha \neq 0$ and C(CP) is explicitly broken then depending
on the value of $\alpha$ the fraction of antimatter in the universe
may be anything from vanishingly small to completely dominant.

\section{Models of anti-baryogenesis \label{sec-anti} }

As follows from the discussion in the previous section the 
necessary conditions for creation of cosmological antimatter
are:
\begin{enumerate}
\item{}
Different sign of C(CP) breaking amplitude in different space 
points. It could be either achieved by spontaneous breaking of
charge symmetry or by a stochastic one created by fluctuations.
\item{}
Exponential but moderate blow-up of regions with a definite sign of
charge symmetry breaking. In the case of too large expansion 
domains of antimatter could exist but far beyond the present day
horizon, while without inflation their size would be too small
and they would mix producing locally charge symmetric universe
with extremely small amount of baryons and antibaryons,
$N_B=N_{\bar B} = 10^{-19} N_\gamma$\cite{chiu66,dolgov81}.
\end{enumerate}

Many models of this kind were considered in the literature,
see e.g.\cite{dolgov92,dolgov93a,chizhov,anti-khlop} and, 
as one can see,
the last two models of the previous section are quite capable
to create a considerable amount of cosmological antimatter. 
In addition to them I would like to 
discuss a rather special one proposed in ref.\cite{dolgov93}.
The model is essentially based of Affleck and Dine scenario
discussed in Sec.~\ref{ssec-susy}. Assume that an additional 
interaction between the field $\chi$ and the inflaton $\Phi$ is 
introduced:
\be
{\cal L}_{int} = \lambda_\Phi |\chi |^2 \left( \Phi - \Phi_1
\right)^2
\label{chi-Phi}
\ee
It is a general renormalizable interaction and its appearance is
quite natural. When the inflaton field is close to some value
$\Phi_1$ the effective mass of $\chi$ reaches minimum and the
transition from one valley to another would be easier. If on the
other hand, the time when $\Phi$ is close to $\Phi_1$ is not very
long then such transition would take place only in a relatively 
small fraction of space. Correspondingly in the dominant part
of the universe volume baryogenesis would create a normal
small baryon asymmetry at the level of $10^{-9}$, while in a small
part of the universe the asymmetry can be very large, even close
to unity. The sign of baryon asymmetry in those high-B domains
may be arbitrary and roughly an equal number of baryonic and
antibaryonic domains with large $|\beta|$ should be created.

According to the calculations made in the paper\cite{dolgov93}
the mass distribution of domains was of log-normal form:
\be
{dN \over dM } \sim \exp \left[ - c \,\ln^2 {M\over M_0} 
\right]
\label{dndm}
\ee
A large part of those (anti)baryon-rich regions would form
primordial black holes. They may be quite heavy with the masses 
up to $10^9 M_{\odot} $, where $M_{\odot} $ is the solar mass.
The very heavy black holes may be the observed today quasars or 
central black holes in large galaxies. But of course not all the
matter (or antimatter) would be hidden inside these black holes 
and a part of it can exist non-collapsed.

About of a half of all such primordial black holes would emerge from
anti-baryon rich regions. Hence some 50\% of central black holes in
galaxies might be made of antimatter. Of course there is no
observational difference between a black hole and an anti black
hole. However, some anti-matter around such black holes may be
non-collapsed and, depending on the amount of the latter, there could
be a flux of radiation from annihilation with the surrounding
matter. It is tempting to explain the switch-off of quasars around
$z=2$ by this mechanism but to this end an optically thick medium
around them is necessary, so the spectrum of annihilation could
degrade down.

In this model clouds of antimatter, or separate stellar objects 
may exist even not too far from us. Especially interesting is that
the primordial nucleosynthesis in this scenario would not stop on
production of light elements, essentially $^4 He$, but because
of a large $\beta$ much more heavy elements could be produced, 
even possibly (anti)iron. This hypothesis may explain observed 
high abundances of metals around quasars at high 
red-shifts\cite{dolgov01} and is supported by the
recent observation\cite{hasinger02} of high iron abundance around
the Quasar APM 08279+5255.  

If we believe the model discussed here, there is a 50\% chance that 
central black hole in our galaxy was made of anti-matter. Moreover,
there should be plenty of lighter ones floating around and not only 
in galactic disk but also in halo, like any other form of cold dark
matter. The parameters of the model can be fixed if we assume that all
dark matter in the universe consists of such primordial black
holes. However, it is not clear if such cold dark matter with a rather
wide mass distribution of the particles (black holes) can give 
a reasonable description of the
observed large scale structure. To answer this question a new type of 
numerical simulation is necessary.

\section{Conclusion}

It seems that creation of astronomically interesting amount of
antimatter is not only possible but quite natural in many
scenarios of baryogenesis. In particular, the conditions for 
creation of antimatter are especially favorable if a condensate of
a complex scalar field existed in the early universe which was
a source of temporary charge parity breaking. 

Theory allows globally charge symmetric universe which, however
may have already problems with the existing observation. On the other
hand, a dominance of matter or antimatter looks equally natural
and a ratio, $\epsilon$, of the amount of cosmic antimatter to 
matter is an unknown parameter of the theory. At the present day
we cannot exclude neither small not large  $\epsilon$ and it 
even possible that we live in a relatively small matter domain in
antimatter dominated universe. Still more detailed and accurate
theoretical calculations in concrete models are necessary since
they could already provide sensible bounds on the possible 
abundance and the distance to the nearest antimatter domains
or astronomical objects. As we discussed in Section~\ref{sec-anti}
antimatter may even live in the halo of our Galaxy.

A search for cosmic antimatter is a very exciting, though extremely
difficult, challenge for the future observations. Its discovery would
indicate to an unusual mechanism of charge symmetry breaking in
cosmology and might be an additional proof of inflationary scenario.
At the moment
it seems that the most promising way to search for cosmic
antimatter is to look for anti-nuclei
in cosmic rays. According to the discussed above models those
could be not only anti-helium but heavier ones up to anti-iron.

\section{Acknowledgments}
I am grateful to Yukawa Institute for Theoretical Physics for
hospitality when the contribution for the Proceedings was prepared.

\end{document}